\documentstyle[12pt,psfig]{article}

\newcommand{\be}{\begin{equation}}
\newcommand{\ee}{\end{equation}}
\newcommand{\ba}{\begin{eqnarray}}
\newcommand{\ea}{\end{eqnarray}}

\begin{document}
\baselineskip=20pt

\begin{center}
  {\Large  \bf Meson resonances from unitarized meson scattering
at one loop in Chiral Perturbation Theory
\footnote{To appear in the proceedings of the 5th International
Conference on ``Quark confinement and the hadron spectrum'', held
in Gargnano, Garda Lake, Italy. 10-14th September 2002}
\footnote{\uppercase{W}ork supported by Spanish CICYT projects
FPA2000-0956, PB98-0782 and BFM2000-1326}}

\vspace{.5cm} {J.R.Pel\'aez$^{1,2}$ 
\footnote{\uppercase{M}arie \uppercase{C}urie \uppercase{F}ellow. 
\uppercase{MCI-2001-01155}. 
\uppercase{E}-mail: jrpelaez@fis.ucm.es },
\lowercase{and} A. G\'omez Nicola$^{2}$}

{$^{1}$ Universita' degli Studi and INFN, Firenze, ITALY\\
 $^{2}$ Dep. F\'{\i}sica Te\'orica II. Universidad Complutense, Madrid. SPAIN
}\vspace{-.5cm}
\end{center}

\noindent
\rule{\textwidth}{.1mm}
\begin{abstract}
\noindent We show the results for the scattering
poles associated to the $\rho$, $f_0$, $a_0$, $K^*$, $\sigma$ 
and $\kappa$ resonances in meson-meson scattering.
Our amplitudes are obtained from the complete one-loop
meson-meson scattering amplitudes from Chiral Perturbation Theory.
Once unitarized with the Inverse Amplitude Method, they 
describe remarkably well the data simultaneously in the
low energy and resonance regions up to 1.2 GeV,
using low energy parameters compatible with present determinations.
\end{abstract}
\vspace{-.5cm}
\rule{\textwidth}{.1mm}

We report on our progress \cite{prep}  on determining the
poles that appear in the recent description \cite{GomezNicola:2001as} of
meson-meson scattering 
by means of unitarized one-loop
Chiral Perturbation Theory (ChPT).
The interest of these poles is that, at least when they are close
to the real axis, they are associated to Breit-Wigner
resonances whose mass $M_R$
and width $\Gamma_R$ is related to the pole position as
$\sqrt{s_{pole}}\simeq M_R-\Gamma_R/2$. When they are not so close
to the real axis, their interpretation is much less clear.

Starting from one-loop Chiral Perturbation Theory \cite{Gasser:1984gg}
our unitarized amplitudes respect the spontaneous chiral symmetry
pattern of QCD up to fourth order in the chiral expansion,
i.e. in powers of meson masses or momenta over the 
chiral scale $4\pi f\simeq 1.2\,$GeV. Nevertheless, since the 
ChPT amplitudes behave as polynomials at high energy, they violate
partial wave unitarity, which we impose with the Inverse
Amplitude Method (IAM).

Part of this program had already been carried first for 
partial waves in the elastic region, for which a single channel
approach could be used, finding the $\rho$ and $\sigma$ poles
in $\pi\pi$ scattering and that of  $K^*$ in $\pi K\rightarrow\pi K$
\cite{Dobado:1996ps}. 
For coupled channel processes, 
an {\it approximate} form of this approach had already been shown
\cite{Oller:1999ag} to yield a remarkable description of meson-meson
scattering up to 1.2 GeV. When these partial waves were continued to
the second Riemann sheet of the complex $s$ several poles were found,
corresponding to the $\rho$, $K^*$, $f_0$, $a_0$, $\sigma$ and $\kappa$
resonances. ( The $\kappa$ pole can also be
obtained in the elastic single channel formalism ).
 The approximations were needed because at that time
not all the meson-meson amplitudes were known to one-loop, and therefore
only the leading order and the dominant s-channel loops were considered
in a simplified calculation, neglecting crossed and tadpole loop
diagrams. As a consequence
the ChPT low energy expansion could only be recovered at leading order.
 The divergences were regularized with a cutoff, which
violates chiral symmetry, making them finite, but not cutoff independent.
Nevertheless, the cutoff dependence was weak and the description
of the data remarkable. Still, due to this cutoff regularization,
it was not possible to compare
 the eight parameters of the chiral Lagrangian,
which are supposed to encode the underlying QCD dynamics, 
with those obtained from other low energy processes. That is, it was
not possible to test the compatibility of the chiral
parameters with the values already present in the literature.

Due to the controversial nature ( or even existence) 
of the scalar states, it is very important to check that
these poles are not just artifacts of the approximations,
estimate the uncertainties in their parameters, and
check their compatibility with other experimental information
regarding ChPT.

The $K\bar{K}\rightarrow K\bar{K}$ amplitudes
were thus calculated in \cite{Guerrero:1998ei}, also unitarizing them
coupled to the $\pi\pi$ states, and reobtaining
the $\sigma$ and $f_0$ and $\rho$ poles.
The whole calculation of meson meson scattering 
has been recently completed with the new
$K\eta\rightarrow K\eta, \eta\eta\rightarrow\eta \eta$ and
$K\eta\rightarrow K\pi$ amplitudes \cite{GomezNicola:2001as}.
In addition the other five existing independent amplitudes
have also been recalculated. The reason is that to 
one loop, one could choose to write all
amplitudes in terms of just  $f_\pi$ (set IAM I\cite{GomezNicola:2001as}), or 
use all $f_\pi$, $f_K$ and $f_\eta$ (set IAM II \cite{prep}), etc... 
However, when one choice is made for one amplitude, 
the other ones have to be calculated consistently in order to
to keep perturbative unitarity, which is needed for the IAM.

Thus, with our recently completed one-loop meson-meson
calculation \cite{GomezNicola:2001as} within the standard
$\overline{MS}-1$ scheme, we have been able to check that
it is possible to find a simultaneous remarkable description of 
meson-meson scattering up to 1.2 GeV, including both the low energy 
and resonance regions. 
We have also been able to estimate the uncertainties
in a fit to the whole meson-meson scattering. Furthermore, we
have shown that this description can be obtained with a set 
of renormalized chiral parameters compatible with those already present
in the literature (see Table I). The new amplitudes reproduce the low
energy chiral expansion up to one loop, in
a remarkable agreement with recent data on threshold
parameters.

\begin{table}[htbp]
{
\small
\setlength{\tabcolsep}{.6mm}
\begin{tabular}{ccccccccc}
($\mu=M_\rho$)&$L_1^r$&$L_2^r$&$L_3$&$L_4^r$&$L_5^r$&$L_6^r$&$L_7$&$L_8^r$\\ \hline
ChPT&0.4$\pm$0.3&1.35$\pm$0.3&-3.5$\pm$1.1&-0.3$\pm$0.5&1.4$\pm$0.5&-0.2$\pm$0.3&-0.4$\pm$0.2&0.9$\pm$0.3\\
IAM I&0.56$\pm$0.10&1.2$\pm$0.1&-2.79$\pm$0.14&-0.36$\pm$0.17&1.4$\pm$0.5&0.07$\pm$0.08&-0.44$\pm$0.15&0.78$\pm$0.18\\
IAM II&0.59$\pm$0.08&1.18$\pm$0.10&-2.93$\pm$0.10&0.2$\pm$0.004&1.8$\pm$0.08&0.0$\pm$0.5&-0.12$\pm$0.16&0.78$\pm$0.7\\\hline
\end{tabular}
}  

{\small Table I. Chiral parameters ($\times10^3$) obtained from IAM
    fits versus previous determinations at $O(p^4)$\cite{Gasser:1984gg,Bijnens:1994ie}.}
\end{table}

Once we have checked the correct chiral low energy behavior,
the scale invariance, and the compatibility of the parameters obtained
from the complete IAM fit, we have very recently extended our
amplitude to the complex plane in search of poles. Our results 
can be found in Table II.

Let us note that we find very stable results for all poles, 
with the exception of the $a_0$, which is very sensible to 
whether one chooses to truncate the series in terms
of a single $f_\pi$ , or also in terms of $f_K$ and $f_\eta$.
We can therefore conclude that the existence of those poles and
their positions are robust results from the Inverse Amplitude Method
when applied to one-loop ChPT meson-meson amplitudes.

\begin{table}[htbp]
{
\begin{tabular}{ccccccc}
\hline
$\sqrt{s_{pole}}$(MeV)
&$\rho$
&$K^*$
&$\sigma$
&$f_0$
&$a_0$
&$\kappa$
\\ \hline
IAM I
&760-i\,82
&886-i\,21
&443-i\,217
&988-i\,4
& cusp?
&750-i\,226
\\
(errors)
&$\pm$ 52$\pm$ i\,25
&$\pm$ 50$\pm$ i\,8
&$\pm$ 17$\pm$ i\,12
&$\pm$ 19$\pm$ i\,3
&
&$\pm$18$\pm$i\,11
\\ \hline
IAM II
&754-i\,74
&889-i\,24
&440-i\,212
&973-i\,11
&1117-i\,12
&753-i\,235
\\
(errors)
&$\pm$ 18$\pm$ i\,10
&$\pm$ 13$\pm$ i\,4
&$\pm$ 8$\pm$ i\,15
&$^{+39}_{-127}$ $^{+i\,189}_{-i\,11}$
&$^{+24}_{-320}$ $^{+i\,43}_{-i\,12}$
&$\pm$ 52$\pm$ i\,33\\\hline
\end{tabular}
}  

{\small Table II. Pole positions (with errors) in meson-meson scattering.
When close to the real axis the mass and width of the 
associated resonance is $\sqrt{s_{pole}}\simeq M-i \Gamma/2$.}
\end{table}


\end{document}